\documentclass[aps,prl,preprint,groupedaddress]{revtex4}

\begin{document}


\title{Quantal Density Functional Theory of Degenerate States}


\author{Viraht Sahni and Xiao-Yin Pan}
\affiliation{Department of Physics, Brooklyn College and the
Graduate School of the City University of New York, New York, New
York 10016. }


\date{\today}

\begin{abstract}
The treatment of degenerate states within Kohn-Sham density
functional theory (KS-DFT) is a problem of longstanding and
 current interest. We propose a solution to this mapping from the
interacting degenerate system to that of the noninteracting
fermion model whereby the  equivalent density and energy are
obtained via the unifying physical framework of quantal density
functional theory (Q-DFT).  We describe the Q-DFT of \textit{both}
\emph{ground} and \emph{excited }degenerate states, and for the
cases of \textit{both} \emph{pure} state and \emph{ensemble}
v-representable densities. The Q-DFT description further provides
a rigorous physical interpretation of the corresponding KS-DFT
energy functionals of the  density, ensemble density, bidensity
and ensemble bidensity , and of their respective functional
derivatives.  We conclude with examples of the mappings within
Q-DFT.

\end{abstract}


\maketitle

The treatment of degenerate states within the context of Kohn-Sham density
functional theory KS-DFT \cite{1} is a problem of longstanding \cite{2} and
 continued recent \cite{3,4} interest.  The basic idea underlying the KS-DFT methodology
  is the mapping from the interacting electronic system as described by Schr{\"o}dinger
  theory to that of a system of \emph{noninteracting fermions} such that the  equivalent
   density and total energy are obtained.  The existence of the model system is an assumption.
 Further, there is the distinction between the S system of
 noninteracting fermions whereby the equivalent density is
 obtained from a \emph{single} Slater determinant, and the
 noninteracting system whose orbitals could be degenerate so that
 the density is obtained from a weighted sum of the Slater
 determinants constructed from the orbitals. The following cases have been considered.  The mapping from:
    (a) a \textit{pure} degenerate ground state\cite{5};
    (b) a \textit{pure} degenerate excited state\cite{2}.
    In addition, maps to obtain the density and energy constructed from
    (c) an \textit{ensemble} of pure degenerate ground states\cite{4}, and
    (d) an \textit{ensemble} of pure degenerate excited states\cite{3}, have been developed.
      The interest in the ensemble cases stems from a \textit{ground} state theorem due to
       Levy \cite{6} and Lieb \cite{7}.  According to the theorem, most ensemble densities
       constructed from pure degenerate ground states are not \textit{interacting}
       v-representable.  In other words, no \textit {single} ground state wave
       function of the Schr{\" o}dinger Hamiltonian will yield this ensemble density.
        Such ensemble densities are said to be \textit{ensemble} v-representable.
         The translation of the theorem to the S system \cite{6} means that there is
         no \textit{single} Slater determinant that leads to this ensemble density.
          At this time, the question of interacting v-representability of the ensemble
          density of degenerate excited states is still unanswered \cite{8}.
          In this paper we propose a solution to the problem of mapping from
          the interacting degenerate system to that of the equivalent noninteracting
          fermion model within the unifying physical framework of quantal density functional
           theory (Q-DFT). \\

In contrast to the present work, the KS-DFT description of the noninteracting system is in
 terms of an energy functional of the density, and its functional derivative.  For case
  (a), the energy is a functional of the degenerate ground state density; for (b),
  the energy is a \textit{bidensity} functional of the ground  and excited state densities,
   with the functional derivative taken at the excited state density; for (c), the energy,
    which is a functional of the ground state ensemble density, is constructed by the
     ensemble generalization of the coupling constant scheme; and for (d) the energy
      is also a \textit{bidensity} functional, in this instance of the ground state
       and excited state ensemble densities, with the functional derivative taken at
        the ensemble density. \\

We describe here the Q-DFT of \textit{both ground} and
\emph{excited} degenerate states, and for the cases of
\textit{both} the \emph{pure} state
 and \emph{ensemble} v-representable densities.  Q-DFT is an alternate
 description \cite{9,10} of the mapping from the interacting to the noninteracting system.
  A key point to note is that within Q-DFT the densities concerned are always
  obtained from solution of the Schr{\"o}dinger equation. Thus,
  the pure state density, and \emph{each} component of the
  ensemble density, are interacting v-representable. The
  assumption of existence of an S system in Q-DFT therefore means that the
  pure state density and \emph{each} component of the ensemble density
  are also noninteracting v-representable. (By an S system we mean a noninteracting fermion system whose wavefunction is a single Slater determinant, and which maybe in a ground or excited state). In Q-DFT, the local electron interaction potential energy $v_{ee}({\bf r})$
of the   S system, and the expression for the total energy are in
terms of `classical' fields
    whose quantal sources are quantum-mechanical expectations of Hermitian operators.
     We begin by (\textit{\textbf{i}}) describing the Q-DFT of the \textit{individual}
     degenerate pure state. For the mapping from a degenerate \textit{ground} state
     of the interacting system, the corresponding S system is in its \textit{ground}
      state.  For the mapping from a degenerate \textit{excited} state, the state of
      the S system is \textit{arbitrary} in that it may be in a \textit{ground} or
      \textit{excited} state.  In either case, the highest occupied eigenvalue is the
       negative of the ionization potential.
       (The mapping from a nondegenerate excited state \cite{10} is similar.)
       For the ground \textit{and }excited state ensemble cases,
       we describe \textit{two} different schemes within Q-DFT.  Thus, (\textit{\textbf{ii}}) in the first, the corresponding
noninteracting  system ensemble density is obtained by
constructing $g$ S systems,
       where $g$ is the degeneracy of the state.  Once again for excited states,
       the $g$ S systems may either be in a ground or excited state or a combination
       of the two.  Next, (\textit{\textbf{iii}}) we describe the Q-DFT  whereby the
       ensemble density is obtained from a \textit{single} noninteracting fermion system whose
       orbitals could be degenerate. The construction
       of this model system is a consequence of the linearity of the differential
        virial theorem. Here the highest occupied eigenvalue is degenerate,
        and the ensemble density is obtained from the resulting Slater determinants as
         described by Ullrich and Kohn \cite{4} whose work in turn is based on that
         of Chayes et al \cite{11}.  Again for the mapping from an excited state,
         the noninteracting system may be in a ground or excited state.  The Q-DFT description
          then (\textit{\textbf{iv}}) provides the physics underlying \textit{all}
           the various KS-DFT degenerate state energy density and bidensity functionals,
            and of their functional derivatives.  Finally, (\textit{\textbf{v}})
            we present examples that demonstrate the above mappings within Q-DFT.\\

  \textbf{A}. The Q-DFT of the bound \textit{individual} degenerate pure state is as
  follows. The Schrodinger equation for a degenerate state whether
  ground or excited is
\begin{equation}
  \hat{H} \Psi_{n,\eta} ({\bf X})=[\hat{T}+ \hat{V}+ \hat{U}]\Psi_{n,\eta}
  ({\bf X})=E_{n} \Psi_{n,\eta} ({\bf X}),
\end{equation}
 $\hat{T}=-\frac{1}{2}\sum_{i}
 {\nabla_{i}}^2$, $\hat{V}=\sum_{i}v({\bf r}_{i})$,
 $ \hat{U}=\frac{1}{2}\sum_{i,j}^{'}\frac{1}{ \mid {\bf r}_{i}-{\bf r}_{j}\mid}
 $, where $\Psi_{n,\eta} ({\bf X})$ and $E_{n}$ are a bound
 degenerate state wave function and energy, n corresponds to the state, and
$\eta =1,...g_{n}$ the degeneracy, ${\bf X}={\bf x_{1}},...{\bf
x_{N}},{\bf x}={\bf r} \sigma$, with $\sigma$ the spin coordinate.
As the equations to follow are valid for arbitrary states, we drop
the subscript n. The degenerate pure state density $\rho_{\eta}
({\bf r})=\langle \Psi_{\eta} \mid \hat{\rho} \mid
\Psi_{\eta}\rangle$, where $\hat{\rho}=\sum_{i}\delta ({\bf
r}-{\bf r}_{i})$, and the energy $E_{\eta}=\langle \Psi_{\eta}
\mid \hat{H} \mid
\Psi_{\eta}\rangle $.\\

  The corresponding differential equation for the S system of
noninteracting fermions with the same density  is
\begin{equation}
 [-\frac{1}{2} {\nabla}^2 + v({\bf r}) + v_{ee,\eta}({\bf r})]
 \phi_{i}({\bf x})=\varepsilon_{i} \phi_{i}({\bf x});   \;\;    i=1,...N,
\end{equation}
with
\begin{equation}
\rho_{\eta} ({\bf r})=\langle \Phi_{\eta} \{\phi_{i}\} \mid
\hat{\rho} \mid \Phi_{\eta}\{\phi_{i}\}\rangle =\sum_{i,\sigma}
\mid \phi_{i}({\bf x})\mid ^2 ,
\end{equation}
and $\Phi_{\eta}\{\phi_{i}\}$ is the \textit{single} Slater determinant of
the orbitals $\phi_{i}({\bf x})$. This is the S system wave function. The electron-interaction
potential energy $v_{ee,\eta}({\bf r})$ is representative of
electron correlations due to the Pauli exclusion principle,
Coulomb repulsion, and Correlation-Kinetic effects.
Correlation-Kinetic contributions to the potential energy are a
consequence of the difference in kinetic energy between the
interacting and noninteracting  systems. The potential energy
$v_{ee,\eta}(\bf{r})$ is the work done  to move a  model Fermion
in the force of a conservative field ${\bf \mathcal{F}}_{\eta}
({\bf r})$:
\begin{equation}
v_{ee,\eta}({\bf r})= -\int _{\infty}^{\bf r} {\bf
\mathcal{F}}_{\eta} ({\bf r'}) \cdot d {\bf l'},
\end{equation}
where ${\bf \mathcal{F}}_{\eta} ({\bf r})={\bf
\mathcal{E}}_{ee,\eta} ({\bf r})+{\bf \mathcal{Z}}_{t_{c},
\eta}({\bf r})$. The
 fields ${\bf \mathcal{E}}_{ee,\eta} ({\bf r})$ and ${\bf \mathcal{Z}}_{t_{c}, \eta} ({\bf r})$
are not necessarily conservative. Their sum always is. The
electron-interaction field $\bf{\mathcal{E}_{ee,\eta}}(\bf{r})$ is
representative of Pauli and Coulomb
 correlation: ${\bf \mathcal{E}}_{ee,\eta} ({\bf r})={\bf e}_{ee,\eta} ({\bf r})/ \rho_{\eta}({\bf r})$, where the
 electron-interaction `force' ${\bf e}_{ee,\eta} ({\bf r})$ is obtained via Coulomb's law as ${\bf e}_{ee,
 \eta} ({\bf r})=\int d {\bf r'}  P_{\eta}({\bf r} {\bf r'})
 ({\bf r}-{\bf r'})/\mid {\bf r}-{\bf r'}\mid ^3$, where $P_{\eta}({\bf r} {\bf
 r'})=\langle \Psi_{\eta} |\hat{P}({\bf r} {\bf
 r'})|\Psi_{\eta}\rangle$, with $\hat{P}({\bf r} {\bf
 r'})=\sum_{i,j}' \delta ({\bf r}-{\bf r}_{i}) \delta ({\bf r'}-{\bf
 r}_{j})$.  Equivalently, the field ${\bf \mathcal{E}}_{ee,
 \eta} ({\bf r})$ may be thought of as being due to its quantal
 source, the pair-correlation density $g_{
 \eta} ({\bf r} {\bf r'})=P_{\eta}({\bf r} {\bf r'})/\rho_{\eta}({\bf r})$. The
 Correlation-Kinetic field  ${\bf \mathcal{Z}}_{t_{c}, \eta}({\bf
 r})= {\bf \mathcal{Z}}_{s, \eta}({\bf r})-{\bf \mathcal{Z}}_{ \eta }({\bf
 r}), \;\; {\bf \mathcal{Z}}_{ \eta }({\bf r})={\bf z}_{ \eta }({\bf
 r};  [\gamma_{\eta}])/\rho_{\eta}({\bf r}), \;  {\bf \mathcal{Z}}_{s, \eta}({\bf r})= {\bf z }_{s, \eta }({\bf
 r};  [\gamma_{s,\eta}])/\rho_{\eta}({\bf r})$, and where ${\bf \mathcal{Z}}_{ \eta }({\bf
 r})$ and ${\bf \mathcal{Z }}_{s, \eta }({\bf r})$ are the interacting and  S
 system kinetic fields, respectively. The kinetic `force' ${\bf z}_{ \eta }({\bf
 r})$ is defined by its component $z_{\eta,\alpha }= 2 \sum _{\beta}
 \partial t_{\alpha, \beta} /\partial r_{\beta}$, with
 $t_{\alpha,\beta}({\bf r},[\gamma _{\eta}])= \frac{1}{4}[
 \partial^2 /\partial r'_{\alpha} \partial r''_{\beta}+\partial^2 /\partial r'_{\beta} \partial
 r''_{\alpha}] \gamma_{\eta}({\bf r'}, {\bf r''})|_{{\bf r'}={\bf
 r''}={\bf r}}$ the kinetic energy density tensor. The source of
 the kinetic field ${\bf \mathcal{Z}}_{ \eta }({\bf r})$ is the
 spinless single particle density matrix $\gamma _{\eta}({\bf r},
 {\bf r'})=\langle \Psi_{\eta} |\hat{X}|\Psi_{\eta}\rangle , \;
   \hat{X}=\hat{A} + i \hat{B}, \; \hat{A}= \frac{1}{2}\sum _{j}
 [\delta({\bf r}_{j}-{\bf r}) T_{j}({\bf a}) + \delta({\bf r}_{j}-{\bf r'})T_{j}(-{\bf a}
 )], \; \hat{B}=-\frac{i}{2}\sum _{j}
 [\delta({\bf r}_{j}-{\bf r}) T_{j}({\bf a}) - \delta({\bf r}_{j}-{\bf r'})T_{j}(-{\bf a}
 )]$, $T_{j}({\bf a})$ is a translation operator, and ${\bf a}={\bf
 r'}-{\bf r}$. The field ${\bf \mathcal{Z}}_{s}({\bf r})$ is
 defined in a similar manner in terms of the S system Dirac
 density matrix $\gamma_{s,\eta} ({\bf r},{\bf r'})=\langle
 \Phi_{\eta}\{\phi_{i}\}|\hat{X}|\Phi_{\eta}\{\phi_{i}\}\rangle=\sum_{i,\sigma}
 \phi_{i}^*( {\bf r} \sigma)\phi_{i}( {\bf r'} \sigma)$.\\

   The proof of Eq.(4) follows by equating the differential
 virial theorems \cite{10,12} for the interacting and S systems
 which are, respectively
\begin{equation}
{\bf \nabla} v({\bf r})= - {\bf F}_{\eta}({\bf r})   \;\;      and
\;\;{\bf \nabla} v({\bf r})=- {\bf F}_{s,\eta}({\bf r}),
\end{equation}
where ${\bf F}_{\eta}({\bf r}) = - {\bf \mathcal{E}}_{ee,
 \eta}({\bf r})+ {\bf \mathcal{D}}_{\eta}({\bf r})+{\bf \mathcal{Z}}_{\eta} ({\bf
 r}), \; {\bf F}_{s,\eta}({\bf r})= {\bf \nabla} v_{ee, \eta}({\bf
 r})+ {\bf \mathcal{D}}_{\eta}({\bf r})+ {\bf \mathcal{Z}}_{s,\eta} ({\bf
 r})$, the differential density field ${\bf \mathcal{D}}_{\eta}({\bf
 r})={\bf d}_{\eta}({\bf r}) /\rho _{\eta}({\bf r}), \; {\bf d}_{\eta}({\bf
 r})= -\frac{1}{4} \nabla \nabla ^2 \rho _{\eta}({\bf r})$. Thus, one
 obtains
\begin{equation}
{\bf \nabla} v_{ee, \eta}({\bf r})=- {\bf \mathcal{F}}_{\eta}({\bf r}),
\end{equation}
from which the interpretation of Eq. (4) follows.

 The
 total energy of the degenerate state  $\eta$ is then
\begin{equation}
E_{\eta}=T_{s,\eta}+ \int \rho _{\eta}({\bf r}) v({\bf r}) d {\bf
r}+ E_{ee,\eta}+ T_{c,\eta},
\end{equation}
where $T_{s,\eta}=\langle
 \Phi_{\eta}\{\phi_{i}\}|\hat{T}|\Phi_{\eta}\{\phi_{i}\}\rangle$
 is the S system kinetic energy, and the electron-interaction $
E_{ee,\eta}$ and Correlation-Kinetic $T_{c,\eta}$ energies in
terms of the respective fields are
\begin{equation}
E_{ee,\eta} = \int d {\bf r} \rho_{\eta}( {\bf r}) {\bf r} \cdot
 {\bf \mathcal{E}}_{ee,\eta}({\bf r}),  \;and
\end{equation}
\begin{equation}
T_{c,\eta}= \frac{1}{2} \int d {\bf r} \rho_{\eta}( {\bf r}) {\bf
r} \cdot {\bf \mathcal{Z}}_{t_{c}, \eta} ({\bf r}).
\end{equation}
 These expressions are
 independent of
whether the fields ${\bf \mathcal{E}}_{ee, \eta}({\bf r})$ and
${\bf \mathcal{Z}}_{t_{c}, \eta} ({\bf r})$ are conservative or
not.\\

 The  S system whereby  the density and total energy equivalent to that
 of the interacting system degenerate state $\eta$  is defined by Eqs.(2)-(4) and
 (7). If the degenerate state is excited , the  S system may be
 constructed  to be either in a ground or excited state. Since the electron-interaction field remains unchanged, the
 difference between the corresponding potential energies is
 independent of the Pauli principle and Coulomb repulsion and
 due entirely to the corresponding Correlation-Kinetic fields ${\bf \mathcal{Z}}_{t_{c}, \eta} ({\bf
 r})$. Hence, the potential energy $v_{ee,\eta}({\bf r})$ is different depending on whether
 the S system is in a ground or excited state. In either case, the highest occupied eigenvalue of the S
 system differential equation is the negative of the ionization
 potential. This follows by equating the asymptotic structure of
 the density for the interacting and S systems.\\

    In the transformation from an excited pure degenerate state to an S system in its ground
    state, the fact that the interacting system wave function has nodes is of no relevance.
    By construction, the S and interacting system density $\rho ({\bf r})$ are equivalent, and the density $\rho({\bf r})\geq 0$. Such a mapping for an excited pure nondegenerate state has been demonstrated in Ref. [10].  \\

 \textbf{B}. We next describe the first of \textit{two} ways of obtaining the
 \textit{ensemble} density and energy of the degenerate states via Q-DFT.
 The interacting system ensemble density matrix  operator $\hat{D}({\bf X}{\bf
 X'})$ is defined as
\begin{equation}
\hat{D}({\bf X}{\bf X'})= \sum_{\eta =1}^g \omega_{\eta}
\Psi^*_{\eta}({\bf X}) \Psi_{\eta}({\bf X'}); \;\; \sum_{\eta
=1}^g \omega_{\eta} =1; \;\; 0 \leq \omega_{\eta} \leq 1,
\end{equation}
so that the ensemble density $\rho_{ens}({\bf r})$ and energy
$E_{ens}$ are respectively
\begin{equation}
\rho_{ens}({\bf r})= tr (\hat D \hat \rho )= \sum_{\eta =1}^g
\omega_{\eta} \rho_{\eta} ({\bf r}),
\end{equation}
and
\begin{equation}
E_{ens}= tr (\hat D \hat H )= \sum_{\eta =1}^g \omega_{\eta}
E_{\eta},
\end{equation}
with $\rho_{\eta} ({\bf r})$ and $E_{\eta}$ as defined
previously. (There are ensemble densities that cannot be represented by a single Slater determinant. However, its pure state component density can always be reproduced by an S system).\\

 For \textit{each }degenerate state $\eta$, the density  $\rho_{\eta} ({\bf
 r})$ and energy $E_{\eta}$  can be constructed from an S system as described in
 part A. Thus, the ensemble density and energy of Eqs. (11) and
 (12) may be obtained from \textit{g} S systems. Each S system contributing to
 the ensemble density  may be in a ground or
 excited state.
 Note that the electron-interaction potential energy $v_{ee,\eta}({\bf r})$ for \textit{each}
 of the $g$ S systems will be different. Further, $v_{ee,\eta}({\bf r})$ will be different depending
 on whether the particular S system is in a ground or excited state as explained previously.
 Thus, the ensemble
 density and energy within Q-DFT are obtained by replacing the $\rho_{\eta} ({\bf
 r})$ and  $E_{\eta}$ on the right hand  sides of Eqs. (11) and
 (12) by the corresponding S system equivalents of Eqs. (3) and
 (7), respectively.\\

 \textbf{C}. The ensemble density and energy may also be determined from a
 noninteracting fermion system whose orbitals could be
 degenerate as constructed within Q-DFT. According to Chayes et
 al \cite{11}, the ground state ensemble density may be determined as a unique
 weighted sum of squares  of a finite number \textit{g} of degenerate
 wave functions of this system. The potential energy $v_{ee}({\bf
 r})$ of these noninteracting fermions is then determined  via Q-DFT as follows.
 Rewrite the interacting and  noninteracting system differential virial
 theorems of Eq. (5) as
 \begin{equation}
 \rho_{ens}({\bf r}) {\bf \nabla } v({\bf r})= \sum_{\eta =1}^g
 \omega_{\eta}{\bf f}_{\eta}({\bf r}),
 \end{equation}
 where $ {\bf f}_{\eta}({\bf r})= {\bf e}_{ee, \eta}({\bf r})-{\bf
 d}_{\eta}({\bf r})-{\bf z}_{\eta} ({\bf r})$, and
 \begin{equation}
 \rho_{ens}({\bf r}) {\bf \nabla} v({\bf r})= \sum_{\eta =1}^g
 \omega_{\eta}{\bf f}_{s,\eta}({\bf r}),
 \end{equation}
 where ${\bf f}_{s,\eta}({\bf r})= \rho_{\eta}({\bf r}){\bf \nabla} v_{ee,\eta}({\bf
 r})- {\bf d}_{\eta}({\bf r})- {\bf z}_{s ,\eta} ({\bf r})$.
 Equating Eqs. (13) and (14) leads to
 \begin{equation}
 \sum_{\eta =1}^g \omega_{\eta} \rho_{\eta}({\bf r}) {\bf \nabla} v_{ee,\eta}({\bf
 r})= \sum_{\eta =1}^g \omega_{\eta}  {\bf q}_{\eta}({\bf r}),
\end{equation}
where ${\bf q}_{\eta}({\bf r})= {\bf e}_{ee, \eta}({\bf r})+ {\bf
z}_{t_{c},\eta}({\bf r}) , \;{\bf z}_{t_{c},\eta}({\bf r})={\bf
z}_{s,\eta}({\bf r})- {\bf z}_{\eta}({\bf r})$. Eq. (15) is a
consequence of the linearity of the differential virial theorem
\cite{12}.
 As we require a \textit{single} effective potential energy $v_{s} ({\bf
 r})= v({\bf r}) + v_{ee} ({\bf r})$, we replace  $v_{ee, \eta} ({\bf
 r})$, in Eq. (15) by $v_{ee} ({\bf r})$ to obtain
\begin{equation}
{\bf \nabla} v_{ee}({\bf r})= - {\bf Q}({\bf r}),
\end{equation}
where ${\bf Q}({\bf r})=-(\sum_{\eta =1}^g \omega_{\eta} {\bf q}_{\eta}({\bf r}))/ \rho_{ens} ({\bf r})$, Thus, the
electron-interaction potential energy $v_{ee}({\bf r})$ is the
work done in the conservative field ${\bf Q}({\bf r})$:
\begin{equation}
v_{ee}({\bf r})= -\int_{\infty}^{\bf r}{\bf Q}({\bf r'}) \cdot d
{\bf l'}.
\end{equation}
Note that the components ${\bf  q}_{\eta}({\bf r})$ are
conservative so that ${\bf Q}({\bf r})$ is conservative, and hence
$v_{ee}({\bf r})$ is path independent.\\

For the occupation of orbitals we follow Ullrich-Kohn \cite{4}.
Accordingly, all levels are occupied except the highest $(h) $
which is q- fold degenerate and partially occupied. The number of
the model fermions  in these levels are $N^{h} \leq 2 q$. The
ensemble density which is a weighted sum of the degenerate Slater
determinants is
\begin{equation}
\rho_{ens}({\bf r})=\sum_{i,\sigma}^{N-N^{h}} |\phi_{i}({\bf
x})|^2 + \sum_{i,\sigma}^{q} f_{i} |\phi_{i}^{h}({\bf x};R)|^{2},
\end{equation}
with $ 0\leq f_{i}\leq 1$, and $f_{i}=\sum_{\eta=1}^{g}
\omega_{\eta} \theta_{i,\eta}$ where $\theta_{i,\eta}=1 $ if the
orbital $\phi_{i}^{h}({\bf x};R)$ occurs in the determinant $
\Phi_{\eta}\{\phi_{i}\}$, and 0 otherwise. Here the
$\phi_{i}^{h}({\bf x};R)$ are appropriately rotated (R) orbitals
determined self-consistently together with the $f_{i}$ and the
lower lying orbitals leading to the ensemble density. The ensemble
 energy is obtained from the \textit{g} Slater determinants as in
part B.  Once
again for an excited state ensemble density, the corresponding noninteracting 
system may be in a ground or
excited state.\\

Note that the methodology of construction of the
\textit{g} S systems of part B also follows from Eq. (15). We believe that
it is easier to construct the $g$ S systems of part B than it is to construct
the single noninteracting system of part C. This is because each of the $g$ S systems may be constructed independently.\\

 From the above degenerate state Q-DFT description it is then possible to provide  a
 rigorous physical interpretation for \textit{each} energy functional and functional
 derivative of the corresponding KS-DFT. In each case, the local potential energy
  of the model fermions is the work done in a conservative field. The energy
  in turn may be expressed in terms of the components of this field. Thus, for example,
  the
 KS-DFT degenerate \textit{ground} state electron-interaction
 energy functional $E_{ee}^{KS}[\rho_{ens}]$ \cite{4} of the
 ensemble density is the ensemble sum of the electron-interaction
 $E_{ee,\eta}$ and Correlation-Kinetic $T_{c,\eta}$ energies. The
 functional derivative is the work done to move the model fermion
 in the conservative field ${\bf Q}({\bf r})$. The \textit{same}
 interpretation applies to the\textit{ bidensity } energy
 functional $E^{KS}_{ee}[\rho _{gr}, \rho_{ens}]$ \cite{3}  of degenerate \textit{excited} state KS-DFT,  and of
 its functional derivative.\\

 The Q-DFT mapping from a pure degenerate excited state to an S
 system can be  demonstrated via the first excited
 \textit{triplet} state of the exactly solvable Hooke's atom
 \cite{13}. This atom is comprised of two electrons with a harmonic
 external potential energy $v({\bf r})=\frac{1}{2}\omega r^{2}$. The
 triplet state wave functions are of the form
\begin{equation}
\Psi({\bf r}_{1} {\bf r}_{2})= C_{0} e^{-\omega R^{2}} e^{-\omega
r^{2}/4} [1+ C_{1} \sqrt{\frac{\omega}{2}} r+ C_{2}
(\frac{\omega}{2}) r^2  +  C_{3} (\frac{\omega}{2} )^{3/2} r^3]
Y_{l,m}(\theta,\phi), \;\;\;
\end{equation}
where $Y_{l,m}(\theta,\phi)$ is the spherical harmonic with $l=1,
m=-1, 0, 1, {\bf r}= {\bf r}_{2}-  {\bf r}_{1}, {\bf R}=( {\bf
r}_{2}+  {\bf r}_{1})/2$, and $C_{0}, C_{1}, C_{2},C_{3}$ are
constants. For each degenerate wave function, the corresponding S
system, in either a ground or excited state, can be determined in
a manner similar to the transformation of the first excited
\textit{singlet} state \cite{10}. These results will be presented
elsewhere. The ensemble density of these degenerate states then
follows from the 3  S systems.
 (Note that the ensemble density for this two electron model is
 v-representable. However, the methodology for constructing the \textit{g}
 S systems is the same whether or not the density is
 v-representable.) Another example is that of the noninteracting Be
 atom \cite{4}. Here the ensemble density, which is not
 v-representable, is the weighted sum of the density of $4$ S
 systems in the states $1 s^{2}2 s^{2}, 1s^{2} 2 p_{i}^2 \;
 (i=x,y,x).$ This latter model atom is also an example \cite{4} of
 the noninteracting fermion system that leads to the ensemble density
 with appropriately rotated highest occupied orbitals. \\

 In conclusion,   we have described via Q-DFT the physics of mapping
from a degenerate state of Schr{\"o}dinger theory to that of a
model system of noninteracting fermions such that the equivalent density
and energy are determined. The cases of \textit{both} \textit{pure} and \textit{ensemble}
 v-representable densities are explained. The framework is general
and formally the \textit{same} for \textit{both}
 degenerate \textit{ground} and \textit{excited} states. \\

 This work was supported in part by the Research Foundation of
 CUNY. \pagebreak

\subsection{}
\subsubsection{}

\end{document}